\documentclass[aps,PRB,reprint,superscriptaddress,preprintnumbers]{revtex4-2}
\usepackage[T1]{fontenc}
\usepackage{times}
\usepackage{graphicx,color}
\usepackage{amsfonts,amsmath,amssymb,amsbsy}
\usepackage[colorlinks=true, linkcolor=blue, citecolor=blue, urlcolor=blue]{hyperref}
\usepackage{float}

\begin{document}

\title{Universal quantum computation based on Nano-Electro-Mechanical Systems%
}
\author{Motohiko Ezawa}
\affiliation{Department of Applied Physics, The University of Tokyo, 7-3-1 Hongo, Tokyo
113-8656, Japan}
\author{Shun Yasunaga}
\affiliation{Department of Electrical Engineering, University of Tokyo, Hongo 7-3-1,
113-8656, Japan}
\author{Akio Higo}
\affiliation{Department of Electrical Engineering, University of Tokyo, Hongo 7-3-1,
113-8656, Japan}
\author{Tetuya Iizuka}
\affiliation{Department of Electrical Engineering, University of Tokyo, Hongo 7-3-1,
113-8656, Japan}
\author{Yoshio Mita}
\affiliation{Department of Electrical Engineering, University of Tokyo, Hongo 7-3-1,
113-8656, Japan}

\begin{abstract}
We propose to use a buckled plate as a qubit, where a double-well potential
is mechanically produced by pushing the plate from both the sides. The right
and left positions of the plate are assigned to be quantum states $|0\rangle 
$ and $|1\rangle $. Quantum effects emerge when 
the displacement is of the order of picometers, although the size of a buckled plate
is of the order of $1\mu m$. The NOT gate is executed by changing the
buckling force acting on the plate, while the Pauli-Z gate and the
phase-shift gate are executed by applying electric field. A two-qubit phase
shift gate is materialized with the use of an electrostatic potential. They
constitute a set of universal quantum gates. 
An examination of material parameters leads to a feasibility of a NEMS(Nano-Electro-Mechanical System)-based quantum
computer.
\end{abstract}

\maketitle

\textbf{Introduction.} According to Moor's law, elements of integrated
circuits become exponentially small as a function of year. The size will
become the order of nanometers within 10 years, where quantum mechanical
effects are inevitable. For example, the superposition of states and the
entanglement occur, which are absent in classical mechanics. It is
impossible to decrease the size of elements smaller than 1nm, which is a
typical scale of atoms. This is the end of Moore's law. Quantum computation%
\cite{Feynman,DiVi,Nielsen} is a candidate of "More than Moore", which
resolves the limit of Moor's law. It gives an exponential speed up for some
algorithms. The problem is how to materialize a qubit based on actual
materials. Various proposals are made such as superconductors~\cite{Nakamura}%
, photonic systems~\cite{Knill}, quantum dots~\cite{Loss}, trapped ions~\cite%
{Cirac}, and nuclear magnetic resonance~\cite{Vander,Kane}. Recently,
nanoscale-skyrmion-based qubits\cite{Psa,SkBit} and meron-based qubits\cite%
{MeronBit} are also proposed.

Micro-Electro-Mechanical System (MEMS) is one of the basic elements in the
current technology\cite{Mita1,Mita2,Toshiyoshi}. They use electrostatic
energy to induce mechanical motions. If the size becomes of the order of
nanometer, they are called Nano-Electro-Mechanical System (NEMS)\cite%
{Cra,Eki}. It has been demonstrated\cite{Blenco,Poot,Slowik} that quantum
effects emerge in the oscillation modes of a cantilever when its sample size
is of the order of 100nm$\sim $1 $\mu $m but with the displacement being of
the order of picometers. It is described by a quantum harmonic oscillator.
Carbon nanotubes, DNAs or biomolecules are used to compose elements in NEMS.
Quantum effects have also been observed for a buckled beam made of a carbon
nanotube as in the case of a cantilever .

The buckled plate has two stable positions. It can be used as a classical
bit. It has been proposed that an Ising annealing machine is executable by a
series of the buckled plates\cite{TriMEMS}. Its mechanism is based on the
electrostatic potential inducing the Ising interaction between two adjacent
plates.

A buckled plate would also reveal quantum effects when the displacement is
of the order of picometers. In this paper, we propose to use it as a quantum
bit, which is well described by a double-well potential. We then propose how
to construct a set of universal quantum gates based on buckled plate MEMS,
which consists of the phase-shift gate, the Hadamard gate and the CNOT gate.
They are constructed by tuning the tension, applying electric field and
voltage. A merit is that it is not necessary to use external magnetic field. 

\textbf{Buckled plate MEMS. }In the field of MEMS, the bistable structure
has been studied with a typical application to memories\cite{Vang,
Inta,TriMEMS}, where a plate is buckled. It was proposed\cite{TriMEMS} to
use this buckled plate as the classical bit information $1$ ($0$), when it
is buckled rightward (leftward).

We push a plate from both the ends. The position along the $x$ axis is
determined as $x=a$ by minimizing the double-well potential\cite%
{Rincon,TriMEMS} $V_{\text{DW}}(x)$, 
\begin{equation}
V_{\text{DW}}(x)=\lambda (x^{2}-a^{2})^{2}.  \label{DW}
\end{equation}%
Explicit representations of $\lambda $ and $a$ are given in terms of
material parameters in Supplemental Material I.

This buckled plate is an example of a MEMS (NEMS), when its size is of the
order of micrometers (nanometers).

\textbf{Quantum NEMS. }The dynamics of a buckled NEMS is described by the
Schr\"{o}dinger equation 
\begin{equation}
i\hbar \frac{d}{dt}\psi \left( x,t\right) =H\psi \left( x,t\right) ,
\label{Scroe}
\end{equation}%
where the Hamiltonian is 
\begin{equation}
H=-\frac{\hbar ^{2}}{2m}\frac{d^{2}}{dx^{2}}+V_{\text{DW}}(x),
\label{HamilDW}
\end{equation}%
together with the double-well potential (\ref{DW}).  In what follows, we
use $t_{\text{u}}=\left( m^{2}/\hbar \lambda \right) ^{1/3}$ and $x_{\text{u}%
}=\hbar ^{1/3}/\left( m\lambda \right) ^{1/6}$ as the units of time and
space, respectively, about which we explain in Fig.S2 in Supplemental
Material II. 

\textbf{Qubit.} We numerically solve the eigenenergies of the double-well
system\cite{Korsch,Math}, and obtain the energy spectrum as in Fig.\ref%
{FigSearch}(a1). As is well known, it consists of undegenerated levels for
small $a$ ($a/x_{\text{u}}\ll 1$) and two-fold degenerated levels for large 
$a$  ($a/x_{\text{u}}\gg 1$). What is unexpected is a sharp transition of
the spectrum at $a\approx 1.5x_{\text{u}}$ as a function of $a$. We propose
to use the lowest two-fold degenerated states at $a\gtrsim 1.5x_{\text{u}}$
as a qubit. Corresponding wavefunctions are shown in Supplemental Material
III. 

We represent the state $\left\vert 0\right\rangle $ by the wave function $%
\psi _{+}(x)$ localized at the right bottom and the state $\left\vert
1\right\rangle $ by the wave function $\psi _{-}(x)$ localized at the left
bottom. Their degeneracy is resolved for $a\lesssim 1.5x_{\text{u}}$, where
the ground state is well described by the symmetric state $(\psi _{+}+\psi
_{-})/\sqrt{2}$. We propose to use this transition of the level splitting at 
$a\approx 1.5x_{\text{u}}$ for a NOT gate operation.

\begin{figure}[t]
\centerline{\includegraphics[width=0.48\textwidth]{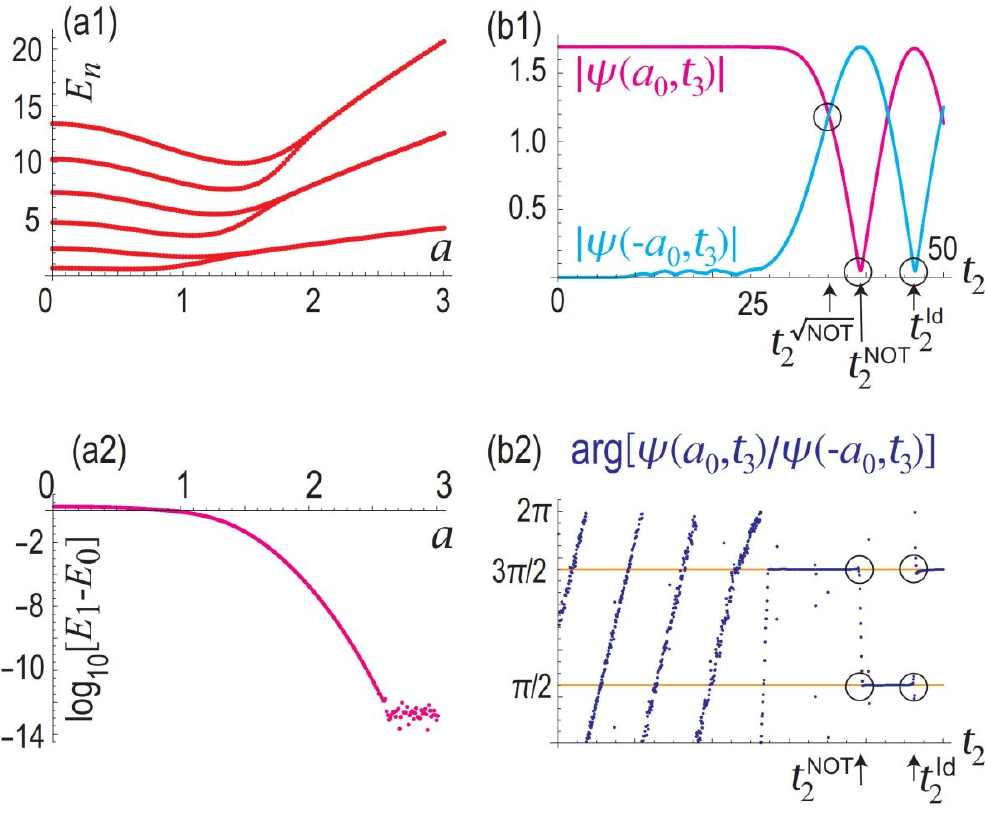}}
\caption{(a1) The energy spectrum as a function of $a$. The lowest six
energy levels are shown. (a2) The logarithm plot $\log _{10}(E_{1}-E_{0})$
of the energy difference between the ground state and the first excited
state. (b1) The absolute value $|\protect\psi (\pm a_{0},t_{3})$ and (b2)
The phase difference arg$\protect\psi (a_{0},t_{3})-$arg$\protect\psi %
(-a_{0},t_{3})$. The horizontal axis is the position $a$ in units of $x_{%
\text{u}}$ in (a1) and (a2), and the time $t$ in units of $t_{\text{u}}$ in
(b1) and (b2). We have set $\mathcal{T}=t_{\text{u}}/5$.}
\label{FigSearch}
\end{figure}

\begin{figure}[t]
\centerline{\includegraphics[width=0.48\textwidth]{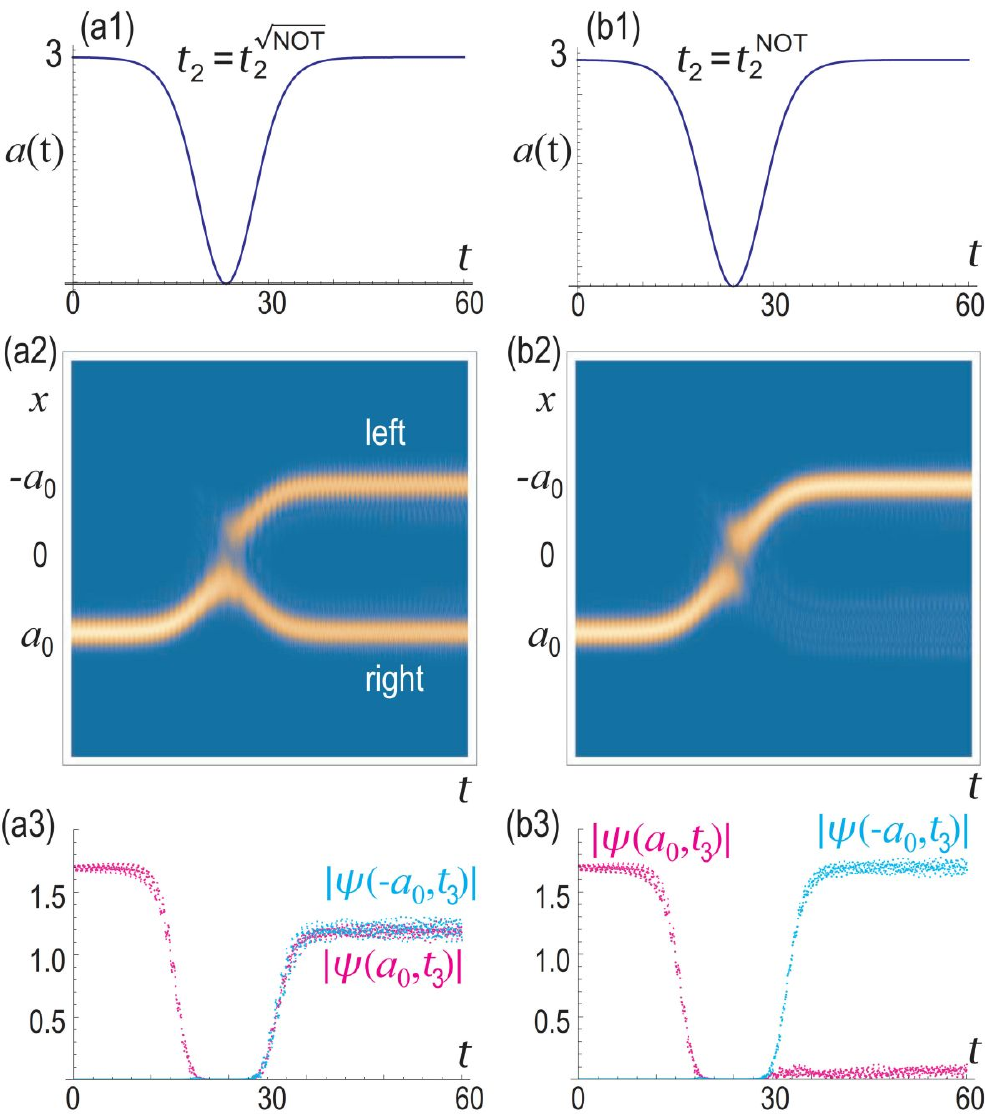}}
\caption{(a) The $\protect\sqrt{\text{NOT}}$ gate, where we have set $t_{2}^{%
\protect\sqrt{\text{NOT}}}=27.02t_{\text{u}}$. (b) The NOT gate, where we
have set $t_{2}^{\text{NOT}}=28.87t_{\text{u}}$. (a1) and (b1) The time
evolution of $a(t)$ given in Eq.(\protect\ref{PosA}) in units of $x_{\text{u}%
}$. (a2) and (b2) The time evolution of the spatial profile of the absolute
value of the wave function $\protect\psi (x,t)$ starting from the localized
state at the right hand side. (a3) and (b3) The time evolution of $|\protect%
\psi (\pm a_{0},t)|$, where $|\protect\psi (a_{0},t)|$ colored in magenta
and $|\protect\psi (-a_{0},t)|$ is colored in cyan. We have set $a_{0}=3x_{%
\text{u}}$, $t_{1}=20t_{\text{u}}$ and $\mathcal{T}=t_{\text{u}}/5$. The
horizontal axis is time ranging $0<t<60t_{\text{u}}$.}
\label{FigPropagate}
\end{figure}

Actually, the two-fold degeneracy is slightly broken for finite $a$. The
energy difference between the ground state and the first-excited state is
calculated. The log$_{10}\left( E_{1}-E_{0}\right) $ is plotted in Fig.\ref%
{FigSearch}(a2). It is found that the energy difference is as tiny as $%
10^{-14}\hbar \sqrt{\lambda /m}$ at $a\approx 2.5x_{\text{u}}$.

The potential is expanded by the harmonic potential 
\begin{equation}
V_{\text{DW}}(x)\simeq 4a^{2}\lambda (x\mp a)^{2}+o\left( (x\mp a)^{3}\right)
\end{equation}%
in the vicinity of $x=\pm a$ with $a\gtrsim 2x_{\text{u}}$, where the
ground-state wave function is given by 
\begin{equation}
\psi _{\pm }(x)=\left( \frac{m\omega }{\hbar \pi }\right) ^{1/4}\exp \left( -%
\frac{m\omega }{2\hbar }(x\mp a)^{2}\right) ,  \label{Psi0}
\end{equation}%
with the characteristic frequency $\omega =2a\sqrt{2\lambda /m}$ and the
ground state energy $E_{0}=\hbar \omega /2$.

\textbf{Universal quantum gates.} It is known that a set of the $\pi /4$
phase-shift gate, the Hadamard gate and the CNOT gate is enough for
constructing any quantum circuits. It is known as the Solovay-Kitaev theorem
of universal quantum computation\cite{Deutsch,Dawson,Universal}. We
explicitly show that they are actually constructed in buckled NEMS.

\textbf{Construction of $\sqrt{\text{NOT}}$ and NOT gates.} Our
scenario reads as follows. Let us start with either the state $\left\vert
0\right\rangle $ or $\left\vert 1\right\rangle $ at $a=a_{0}\gtrsim 2x_{%
\text{u}}$. For definiteness we take $a_{0}=3x_{\text{u}}$. When we change
adiabatically the stable position $a$ from $a_{0}$ to $a=0$, the state is
moved to the symmetric state. Then, we change adiabatically the position $a$
back to the point $a_{0}$. More explicitly, by pushing the plate from the
both ends, we temporally control the stable position $a$ according to a
smooth function,%
\begin{equation}
a\left( t\right) =\frac{a_{0}}{2}\left[ \tanh \frac{t-t_{2}}{\mathcal{T}}%
-\tanh \frac{t-t_{1}}{\mathcal{T}}+2\right] ,  \label{PosA}
\end{equation}%
with three parameters $t_{1}$, $t_{2}$ and $\mathcal{T}$. The resultant
state needs not be the state $\left\vert 0\right\rangle $ or $\left\vert
1\right\rangle $ but can be a combination of $\left\vert 0\right\rangle $
and $\left\vert 1\right\rangle $ in general.

We study the dynamics of the wave packet by numerically solving the Schr\"{o}%
dinger equation (\ref{Scroe}) with a time-dependent Hamiltonian (\ref%
{HamilDW}), where the double-well potential (\ref{DW}) is time-dependent
with the use of the time-dependent position (\ref{PosA}). 

We start from the initial state $\psi _{+}$ given by Eq.(\ref{Psi0})
localized at the right-hand side. In order to see the result of the gate
operation, we focus on the state $\psi \left( x,t\right) $ after enough time
of the gate operation at $t=t_{3}\gg t_{2}$. We show the amplitudes $%
\left\vert \psi \left( \pm a_{0},t_{3}\right) \right\vert $ and the phase
shifts arg$\psi \left( \pm a_{0},t_{3}\right) $ as a function of $t_{2}$ in
Fig.\ref{FigSearch}(b1) and (b2), respectively. We have found that the
amplitude changes as a function of $t_{2}$ significantly. We also find that
the phase difference is $\pi /2$ and $3\pi /2$ shown in Fig.\ref{FigSearch}%
(a2). The jumps occur where the $\left\vert \psi \left( \pm
a_{0},t_{3}\right) \right\vert =0$.

$\sqrt{\text{NOT}}$\textit{gate:} We first construct the square-root NOT
gate, 
\begin{equation}
U_{\sqrt{\text{NOT}}}^{\pm }=\frac{1}{\sqrt{2}}\left( e^{i\pi /4}I_{2}\pm
e^{-i\pi /4}\sigma _{x}\right) ,
\end{equation}%
which satisfies $\left( U_{\sqrt{\text{NOT}}}^{\pm }\right) ^{2}=\pm \sigma
_{x}$. We observe in Fig.\ref{FigSearch}(b1) that the amplitude $|\psi
\left( a_{0},t\right) |$ at the initial position colored in magenta
decreases for $t_{2}>26t_{\text{u}}$, and it becomes identical to $|\psi
\left( -a_{0}\right) |$ colored in cyan at $t_{2}^{\sqrt{\text{NOT}}%
}=27.02t_{\text{u}}$. This value of the parameter is special, where we study
the time evolution of the spatial profile. The result is given in Fig.\ref%
{FigPropagate}(a2), where the wave packet is split equally to the right
and left positions. The time evolution of the amplitude at $x=\pm a_{0}$ is
shown in Fig.\ref{FigPropagate}(a3). We find that the wave function becomes
stationary after the gate operation for $t\gtrsim 40t_{\text{u}}$. Precisely
in the same way, the equal splitting occurs when we start from the initial
state $\psi _{-}$ given by Eq.(\ref{Psi0}) localized at the left-hand side.
This gate operation at $t_{2}^{\sqrt{\text{NOT}}}$ is summarized as $U_{%
\sqrt{\text{NOT}}}^{\pm }$.

\textit{NOT gate: }Next, we construct the NOT gate $U_{\text{NOT}}\equiv
\sigma _{x}$. We observe in Fig.\ref{FigSearch}(b1) that the amplitude $%
|\psi \left( a_{0},t\right) |$ at the initial position colored in magenta
becomes zero at $t_{2}^{\text{NOT}}=28.87t_{\text{u}}$. This value of the
parameter is also special, where the wave packet moves to the left position
as shown in Fig.\ref{FigPropagate}(b2). The corresponding time evolution of
the amplitude at $x=\pm a_{0}$ is shown in Fig.\ref{FigPropagate}(b3). This
is the NOT gate $\sigma _{x}$. We find that the wave function becomes
stationary after the gate operation for $t\gtrsim 40t_{\text{u}}$.

\begin{figure}[t]
\centerline{\includegraphics[width=0.48\textwidth]{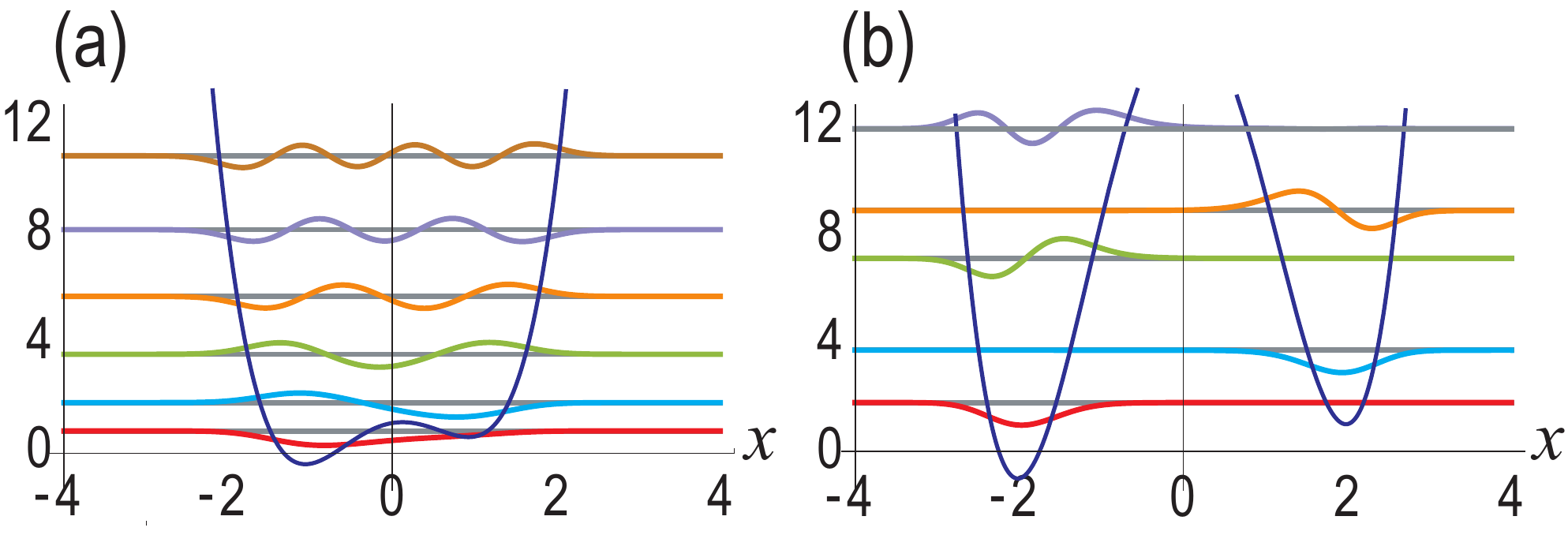}}
\caption{The wave functions and the energy spectrum under electric field.
(a) $a=x_{\text{u}}$ and (b) $a=2x_{\text{u}}$. The ground state wave
function is colored in red, while the first-excited states wave function is
colored in cyan. We have set $E_{x}=0.5$ in units of $E_{\text{u}}\equiv
\hbar \protect\omega /a_{0}$.}
\label{FIgCantWave}
\end{figure}

\begin{figure}[t]
\centerline{\includegraphics[width=0.48\textwidth]{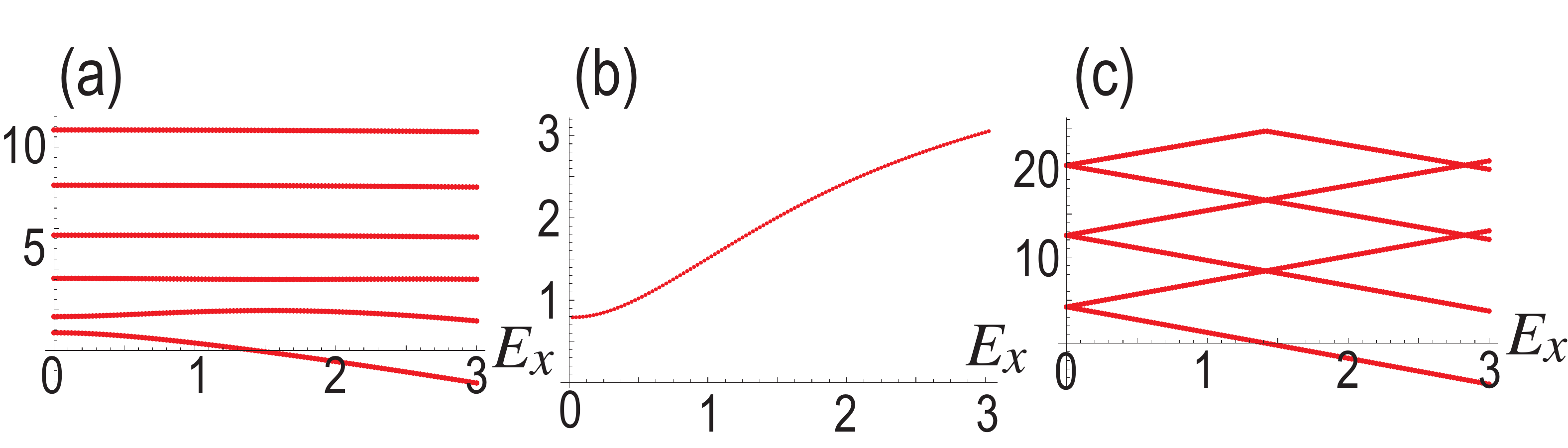}}
\caption{The energy spectrum as a function of the applied electric field $%
E_{x}$ in the case of (a) $a=x_{\text{u}}$ and (c) $a=3x_{\text{u}}$ (b) The
energy difference between the ground state and the first excited state in
the case of $a=x_{\text{u}}$. The horizontal axis is $E_{x}$ in units of $E_{%
\text{u}}\equiv \hbar \protect\omega /a_{0}$}
\label{FIgCantWell}
\end{figure}

\begin{figure}[t]
\centerline{\includegraphics[width=0.48\textwidth]{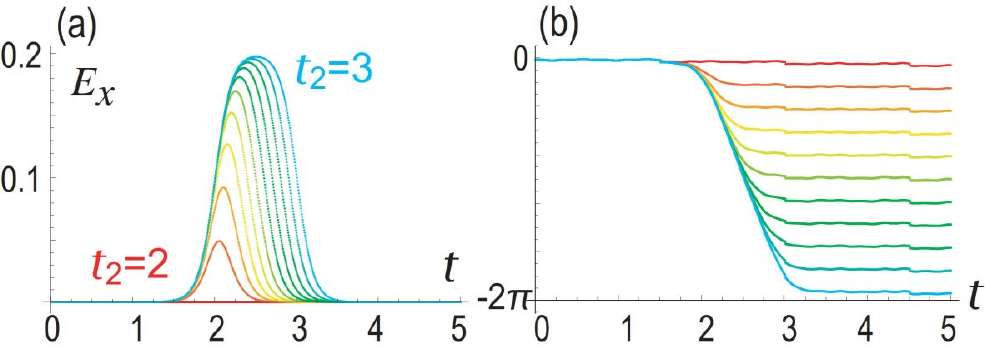}}
\caption{(a) The time dependence of the applied electric field according to (%
\protect\ref{Ext}) $E_{x}$. (b) The time evolution of the phase difference
compared with the phase evolution without electric field. The color
indicates the time $t_{2}$ ranging from $t_{2}=2t_{\text{u}}$ colored in red
to $t_{2}=3t_{\text{u}}$ colored in cyan. We have set $t_{1}=2t_{\text{u}}$.
The horizontal axis is time ranging $0<t<5t_{\text{u}}$. We have set $%
\mathcal{T}=t_{\text{u}}/5$. }
\label{FIgPhaseSolve}
\end{figure}

\textbf{Construction of phase-shift and Pauli-Z gates.} We proceed to
construct the phase-shift gate. We apply an electric field along the $x$
direction to the buckled plate, where the potential is given by $E_{x}x$.
The potential and eigenfunctions under electric field, $V_{\text{DW}%
}(x)+E_{x}x$, are shown in Fig.\ref{FIgCantWave}. We numerically evaluate
the energy spectrum as a function of the electric field, which is shown in
Fig.\ref{FIgCantWell}(a). The energy difference between the ground state and
the first-excited state monotonically increases as the increase of electric
field as shown in Fig.\ref{FIgCantWell}(b). For $a_{0}=3x_{\text{u}}$, the
energy spectrum changes linearly as a function of the electric field as
shown in Fig.\ref{FIgCantWell}(c). In the first-order perturbation theory,
the energy is estimated as%
\begin{equation}
\langle \psi _{\pm }|E_{x}x|\psi _{\pm }\rangle =\pm E_{x}a_{0},  \label{Exa}
\end{equation}%
which is consistent with the numerical results shown in Fig.\ref{FIgCantWell}%
(c). Namely, the energy-shift (\ref{Exa}) is represented by the effective
Hamiltonian $H_{E_{z}}=a_{0}E_{x}\sigma _{z}$.

We have numerically evaluated the dynamics of the wave packet starting from
the Gaussian distribution (\ref{Psi0}) under the temporally controlled
electric field according to the formula%
\begin{equation}
E_{x}\left( t\right) =\frac{E_{0}}{2}\left[ \tanh \frac{t-t_{1}}{\mathcal{T}}%
-\tanh \frac{t-t_{2}}{\mathcal{T}}\right] ,  \label{Ext}
\end{equation}%
as shown in Fig.\ref{FIgPhaseSolve}(a). We found that the absolute value $%
\left\vert \psi \right\vert $does not change. However, the phase is
modulated as in Fig.\ref{FIgPhaseSolve}(b) for the state $|\psi _{+}\rangle $%
. The phase modulation for the state $|\psi _{-}\rangle $ is precisely
opposite of that of the state $|\psi _{+}\rangle $. Hence, the result is
summarized as the unitary operator%
\begin{equation}
U_{Z}\left( \theta \right) =\text{diag.}\left( e^{-i\theta /2},e^{i\theta
/2}\right) =\exp \left[ -\frac{i\theta }{2}\sigma _{z}\right] ,
\label{Zrotation}
\end{equation}%
where $\theta $ is determined as a function of $t_{2}$ as illustrated in
Fig.\ref{FIgPhaseSolve}(b). This is the phase-shift gate by angle $\theta $.

$\pi /4$ \textit{phase-shift gate: }The $\pi /4$ phase-shift gate $%
U_{T}\equiv $diag.$(1,e^{i\pi /4})$ is realized by the $z$ rotation (\ref%
{Zrotation}) with the angle $\theta =\pi /4$ as $U_{T}=e^{-i\pi
/8}U_{Z}\left( \frac{\pi }{4}\right) $ up to the overall phase factor $%
e^{i\pi /8}$.

\textit{Pauli-Z gate:} The Pauli-Z gate is realized by the $z$ rotation with
the angle $\pi $ as $U_{Z}=-iU_{Z}\left( \pi \right) $ in a similar way.

\textbf{The Hadamard gate.} The Hadamard gate is defined by $U_{\text{H}%
}\equiv \left( \sigma _{z}+\sigma _{x}\right) /\sqrt{2}$. It is realized by
a sequential application of the $z$ rotation and the $x$ rotation~\cite%
{Schuch} as%
\begin{equation}
U_{\text{H}}=-iU_{Z}U_{\text{NOT}}U_{Z}.  \label{HZXZ}
\end{equation}

\textbf{Two-qubit phase-shift gate.} Next, we construct two-qubit gates made
of two buckled plates. When we apply the voltage $V_{1}$ between the two
plates, the potential energy is given by%
\begin{equation}
V\left( x_{1},x_{2}\right) \equiv V_{\text{DW}}\left( x_{1}\right) +V_{\text{%
DW}}\left( x_{2}\right) +\frac{C_{\text{para}}(x_{1},x_{2})}{2}V_{1}^{2},
\label{TetraV}
\end{equation}%
with $C_{\text{para}}(x_{1},x_{2})$ the capacitance between the plates,

\begin{equation}
C_{\text{para}}(x_{1},x_{2})=\frac{\varepsilon _{0}S}{X_{\text{cap}%
}+x_{1}-x_{2}},  \label{Cu}
\end{equation}%
where $\varepsilon _{0}$ and $S$ are the permittivity and the plate area,
while $X_{\text{cap}}$ is the distance between the two plates. We assume
that the plate distance $X_{\text{cap}}$ is very large compared with the
displacement $a$. We calculate%
\begin{eqnarray}
V\left( a,a\right) &=&V\left( -a,-a\right) =\frac{\varepsilon _{0}S}{X_{%
\text{cap}}}\frac{V_{1}^{2}}{2}\equiv E_{0}, \\
V\left( a,-a\right) &=&\frac{\varepsilon _{0}S}{X_{\text{cap}}+2a}\frac{%
V_{1}^{2}}{2}\equiv E_{+}, \\
V\left( -a,a\right) &=&\frac{\varepsilon _{0}S}{X_{\text{cap}}-2a}\frac{%
V_{1}^{2}}{2}\equiv E_{-}.
\end{eqnarray}%
Then, the potential differences are given by%
\begin{eqnarray}
E_{+}-E_{0} &\simeq &-\frac{a}{X_{\text{cap}}^{2}}\varepsilon
_{0}SV_{1}^{2}=-E_{X},  \label{Eapprox} \\
E_{-}-E_{0} &\simeq &\frac{a}{X_{\text{cap}}^{2}}\varepsilon
_{0}SV_{1}^{2}=E_{X}.  \label{Eapprox2}
\end{eqnarray}%
The detailed derivation is shown in Supplemental Material IV.

We start with the Gaussian state $\Psi _{\sigma _{1}\sigma _{2}}\left(
x_{1},x_{2}\right) \equiv \psi _{\sigma _{1}}(x_{1})\psi _{\sigma
_{2}}(x_{2})$ with Eq.(\ref{Psi0}) localized at four points $x_{1}=\sigma
_{1}a$ and $x_{2}=\sigma _{2}a$, where $\sigma _{1}=\pm ,\sigma _{2}=\pm $.
The absolute value of this wave function almost remains as it is for a
potential at $a/x_{\text{u}}\gg 1$, but a phase shift occurs. The unitary
evolution is given by 
\begin{equation}
U\left( t\right) =\exp [-i\left( E_{0}/\hbar +\omega \right) t]
\end{equation}%
for $\sigma _{1}=\sigma _{2}=+$ and $\sigma _{1}=\sigma _{2}=-$, 
\begin{equation}
U\left( t\right) =\exp [-i\left( E_{+}/\hbar +\omega \right) t]
\end{equation}%
for $\sigma _{1}=+$ and $\sigma _{2}=-$, 
\begin{equation}
U\left( t\right) =\exp [-i\left( E_{-}/\hbar +\omega \right) t]
\end{equation}%
for $\sigma _{1}=-$ and $\sigma _{2}=+$, where we have added the zero-point
energy.

It corresponds to the two-qubit phase-shift gate%
\begin{eqnarray}
U_{\text{2-phase}}\left( t\right) &=&\text{diag.}\left( e^{-i\frac{E_{0}}{%
\hbar }t},e^{-i\frac{E_{-}}{\hbar }t},e^{-i\frac{E_{+}}{\hbar }t},e^{-i\frac{%
E_{0}}{\hbar }t}\right)  \notag \\
&=&e^{-i\frac{E_{0}}{\hbar }t}\text{diag.}\left( 1,e^{-i\frac{E_{X}}{\hbar }%
t},e^{i\frac{E_{X}}{\hbar }t},1\right) ,
\end{eqnarray}%
by identifying the qubit state $\left( \left\vert 00\right\rangle
,\left\vert 01\right\rangle ,\left\vert 10\right\rangle ,\left\vert
11\right\rangle \right) ^{t}=\left( \left\vert ++\right\rangle ,\left\vert
+-\right\rangle ,\left\vert -+\right\rangle ,\left\vert --\right\rangle
\right) ^{t}$.

\textit{Ising gate:} The Ising gate $U_{ZZ}\equiv $diag.$(1,-1,-1,1)$, by
setting $E_{X}t/\hbar =\pi $ up to the global phase $\exp \left[
-iE_{0}t/\hbar \right] $.

\textit{CZ gate:} The controlled-Z (CZ) gate $U_{\text{CZ}}$ is a unitary
operation acting on two adjacent qubits defined by $U_{\text{CZ}}=$diag.$%
(1,1,1,-1)$. We construct it by a sequential application of the Ising gate
and the one-qubit phase-shift gates as\cite{Mak}%
\begin{equation}
U_{\text{CZ}}=e^{i\pi /4}U_{Z}\left( \frac{\pi }{2}\right) U_{Z}\left( \frac{%
\pi }{2}\right) U_{ZZ},  \label{CZZZ}
\end{equation}%
By using the results (\ref{Eapprox}) and (\ref{Eapprox2}), it is enough to
set $\phi _{1}=\pi -\phi _{2}=E_{X}t/\hbar $ to construct the CZ gate.

\textit{CNOT gate:} The CNOT is constructed by a sequential application of
the CZ gate (\ref{CZZZ}) and the Hadamard gate (\ref{HZXZ}) as $U_{\text{CNOT%
}}^{1\rightarrow 2}=U_{\text{H}}^{\left( 2\right) }U_{\text{CZ}}U_{\text{H}%
}^{\left( 2\right) }$. See the definition of the CNOT gate in Supplemental
Material V.

\textbf{Discussions.} We would like to address the feasibility of a
NEMS-based quantum computer by examining typical material parameters\cite%
{Poot}. The length of an element is of the order of 1$\mu $m$\sim $100 $\mu 
$m. The displacement is of the order of 0.1pm$\sim $10fm. The mass $m$ is
of the order of $10^{-21}$kg$\sim 10^{-14}$kg. The characteristic frequency
is of the order of $\sqrt{\kappa /m}$, or 1MHz$\sim $1GHz. Then, such a
NEMS-based quantum computer would be realizable by improving the present
technology.

M.E. is grateful to N. Nagaosa for helpful discussions on the subject. This
work is supported by CREST, JST (Grants No. JPMJCR20T2). 

\clearpage\newpage
\onecolumngrid
\def\theequation{S\arabic{equation}}
\def\thefigure{S\arabic{figure}}
\def\thesubsection{S\arabic{subsection}}
\setcounter{figure}{0}
\setcounter{equation}{0}

\begin{center}
\textbf{\Large Supplemental Material}
\bigskip
\bigskip

\textbf{\large
Universal quantum computation based on Nano-Electro-Mechanical Systems%
}\bigskip

{Motohiko Ezawa}

{Department of Applied Physics, The University of Tokyo, 7-3-1, Hongo,
Bunkyo-ku, Tokyo 113-8656, Japan}
\medskip

{Shun Yasunaga, Akio Higo, Tetuya Iizuka, Yoshio Mita}

{Department of Electrical Engineering, University of Tokyo, Hongo 7-3-1,
113-8656, Japan}

\end{center}
\bigskip

\section{Buckled plate MEMS}

\ We consider a plate with length $2L_{0}$\ and the spring constant $\kappa $%
\ placed along the $y$\ axis. The form of the buckled plate is determined by
the Euler-Bernoulli equation\cite{Vang,Vang2},%
\begin{equation}
\frac{d^{4}w}{dy^{4}}+m^{2}\frac{d^{2}w}{dy^{2}}=0,
\end{equation}%
with $m^{2}\equiv P/EI$, where $P$ is the axial load, $E$ is Young's modulus
of the beam material and $I$ is the second moment of area of the beam. In
solving the Euler-Bernoulli equation for a clamped-clamped beam, we may use
the fixed boundary condition or the free boundary condition.

First, we impose the fixed boundary condition, which reads%
\begin{equation}
w\left( -y_{0}\right) =w\left( y_{0}\right) =0,\qquad \left. \frac{dw}{dy}%
\right\vert _{x=-y_{0}}=\left. \frac{dw}{dy}\right\vert _{x=y_{0}}=0,
\end{equation}%
where $y_{0}$ is the position of the supporting point along the $y$ axis

\begin{figure}[b]
\centerline{\includegraphics[width=0.92\textwidth]{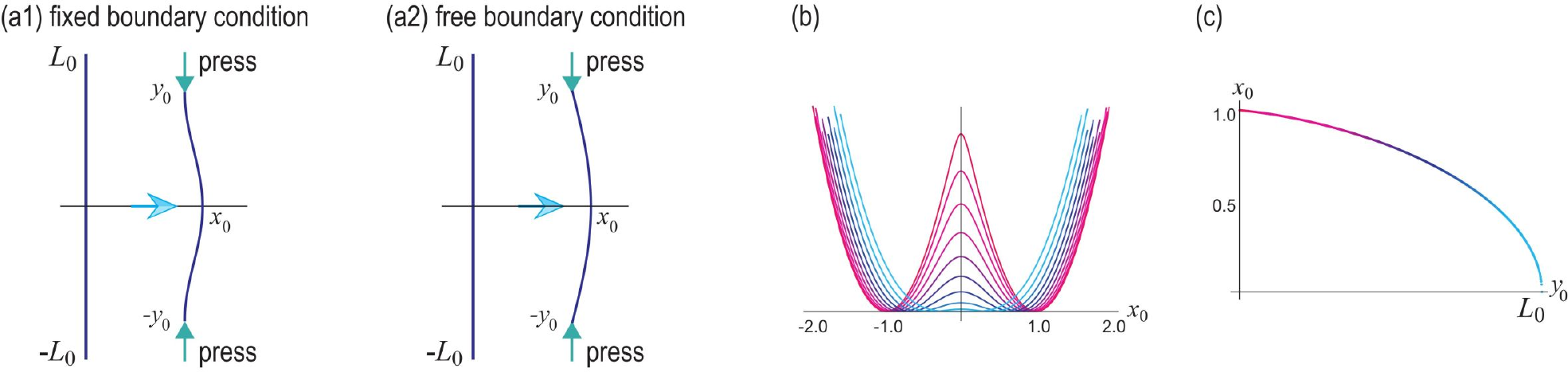}}
\caption{Illustration of a buckled plate with (a1) fixed boundary condition
and (a2) free boundary condition. (b) Electric field is applied along the $x$
direction in order to construct the Pauli-Z gate. (c) The electrostatic
potential is induced by applying voltage in order to construct the two-qubit
phase-shift gate.}
\label{FigPlate}
\end{figure}

The lowest energy solution is given by\cite{Charlot}%
\begin{equation}
w=\frac{x_{0}}{2}\left( 1+\cos \frac{\pi y}{y_{0}}\right) ,
\end{equation}%
where $x_{0}$ is the position along the $x$ axis. The length of the buckled
plate is given by $L(x_{0})=\int_{-y_{0}}^{y_{0}}dy\sqrt{1+(dw/dy)^{2}}$. It
is calculated as%
\begin{eqnarray}
L(x_{0}) &=&\int_{-y_{0}}^{y_{0}}dy\sqrt{1+\frac{\pi ^{2}x_{0}^{2}}{%
4y_{0}^{2}}\sin ^{2}\frac{\pi y}{y_{0}}}=\frac{y_{0}}{\pi }\int_{-\pi }^{\pi
}d\theta \sqrt{1+\frac{\pi ^{2}x_{0}^{2}}{4y_{0}^{2}}\sin ^{2}\theta } 
\notag \\
&=&\frac{4y_{0}}{\pi }\int_{0}^{\pi /2}d\theta \sqrt{1+\frac{\pi
^{2}x_{0}^{2}}{4y_{0}^{2}}\sin ^{2}\theta }=\frac{4y_{0}}{\pi }E\left( -i%
\frac{\pi x_{0}}{2y_{0}}\right) ,  \label{S4}
\end{eqnarray}%
where $E$ is the complete elliptic integral of the second kind defined by%
\begin{equation}
E\left( k\right) \equiv \int_{0}^{\pi /2}\sqrt{1-k^{2}\sin ^{2}\theta }%
d\theta .
\end{equation}%
The potential energy is given by the Hooke law%
\begin{equation}
V(x_{0})=\frac{\kappa }{2}\left( 2L(x_{0})-2L_{0}\right) ^{2},  \label{Vx}
\end{equation}%
which we show for various $y_{0}$ in Fig.\ref{FigPlate}(b). It is a
double-well potential. The potential minimum is shown as a function of $y_{0}
$ in Fig.\ref{FigPlate}(c).

We expand Eq.(\ref{Vx}) in terms of $x_{0}$ by fixing $L_{0}$ and $y_{0}$ as%
\begin{equation}
V(x_{0})=\frac{\kappa }{2}\left( 2L_{0}-2y_{0}\right) ^{2}+\frac{1}{4y_{0}}%
\left( L_{0}-y_{0}\right) \kappa \pi ^{2}x_{0}^{2}+\frac{\left(
L_{0}-3y_{0}\right) }{256y_{0}^{3}}\kappa \pi ^{4}x_{0}^{4}+o\left(
x_{0}^{6}\right) .
\end{equation}%
It is summarized as $V(x_{0})=\lambda (x_{0}^{2}-a^{2})^{2}+V_{0}+o\left(
x_{0}^{6}\right) $ with%
\begin{equation}
\lambda \equiv \frac{\left( L_{0}-3y_{0}\right) }{256y_{0}^{3}}\kappa \pi
^{4},\quad a\equiv y_{0}\frac{4\sqrt{2}}{\pi }\sqrt{\frac{L_{0}-y_{0}}{%
3L_{0}-y_{0}}}.
\end{equation}%
The parameter $x_{0}$ is determined as $x_{0}=a$ by minimizing the potential 
$V(x_{0})$. This buckled plate is an example of a MEMS (NEMS), when its size
is of the order of micrometers (nanometers).

Second, we impose the fixed boundary condition, which reads%
\begin{equation}
\left. \frac{d^{2}w}{dy^{2}}\right\vert _{x=-y_{0}}=\left. \frac{d^{2}w}{%
dy^{2}}\right\vert _{x=y_{0}}=0,
\end{equation}%
whose solution is\cite{Binh} 
\begin{equation}
w=x_{0}\cos \frac{\pi y}{2y_{0}}.
\end{equation}%
The length $L(x_{0})$\ is given by Eq.(\ref{S4}) precisely as in the case of
the fixed boundary condition. Hence, the potential energy is given by the
same Hooke law as Eq.(\ref{Vx}).

\section{Units of time and space}

It is straightforward to rewrite the Schr\"{o}dinger equation (2) and the
Hamiltonian (3) in the main text as 
\begin{equation}
i\frac{d}{d\tau }\psi \left( X,\tau \right) =H\psi \left( X,\tau \right) 
\end{equation}%
and%
\begin{equation}
H=-\frac{1}{2}(d/dX)^{2}+(X^{2}-A^{2})^{2}
\end{equation}%
in terms of the dimensionless time $\tau =t/t_{\text{u}}$ and the
dimensionless parameters $X\equiv x/x_{\text{u}}$ and $A\equiv a/x_{\text{u}}
$, where $t_{\text{u}}=\left( m^{2}/\hbar \lambda \right) ^{1/3}$ and $x_{%
\text{u}}=\hbar ^{1/3}/\left( m\lambda \right) ^{1/6}$ give the units of
time and space, respectively. We have carried out numerical analysis based
on the dimensionless formulas.

\section{Wavefunction}

\begin{figure}[t]
\centerline{\includegraphics[width=0.68\textwidth]{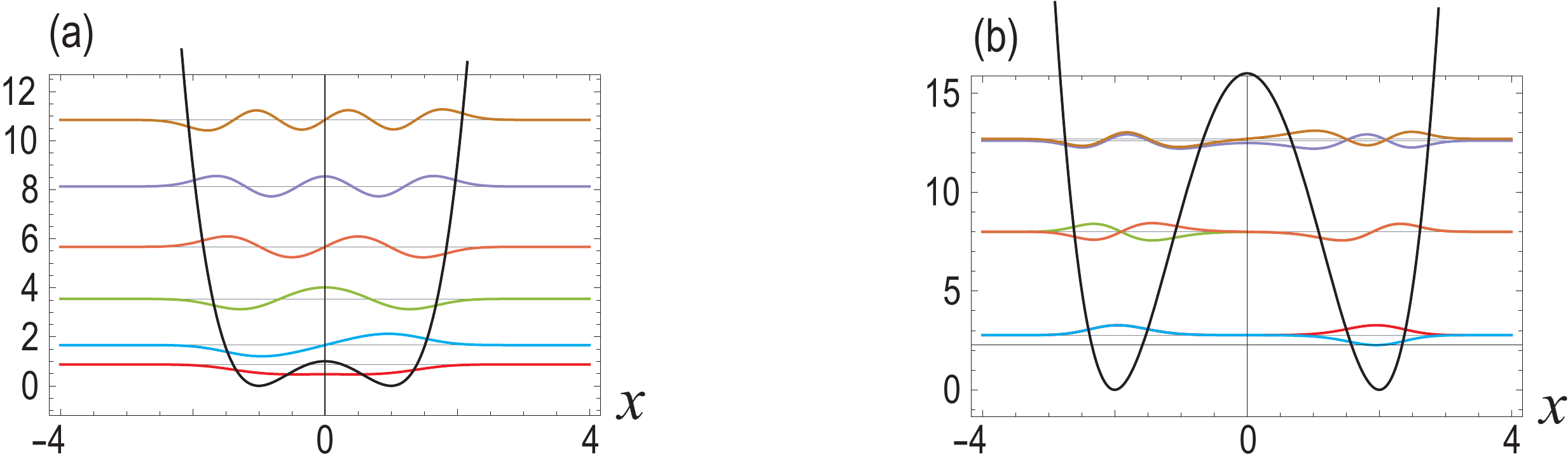}}
\caption{Wave functions and energy spectrum with varying stress. (a) $a=x_{%
\text{u}}$, (b) $a=2x_{\text{u}}$ and (c) $a=3x_{\text{u}}$. The horizontal
axis is $x$. The ground state wave function is colored in red, while the
first-excited states wave function is colored in cyan.}
\label{FIgDoubleWave}
\end{figure}

We also numerically determine the wave functions for a double-well
potential, which are shown in Fig.\ref{FIgDoubleWave}. They are well
described by those of the harmonic potential for large $a$. 

\begin{figure}[t]
\centerline{\includegraphics[width=0.48\textwidth]{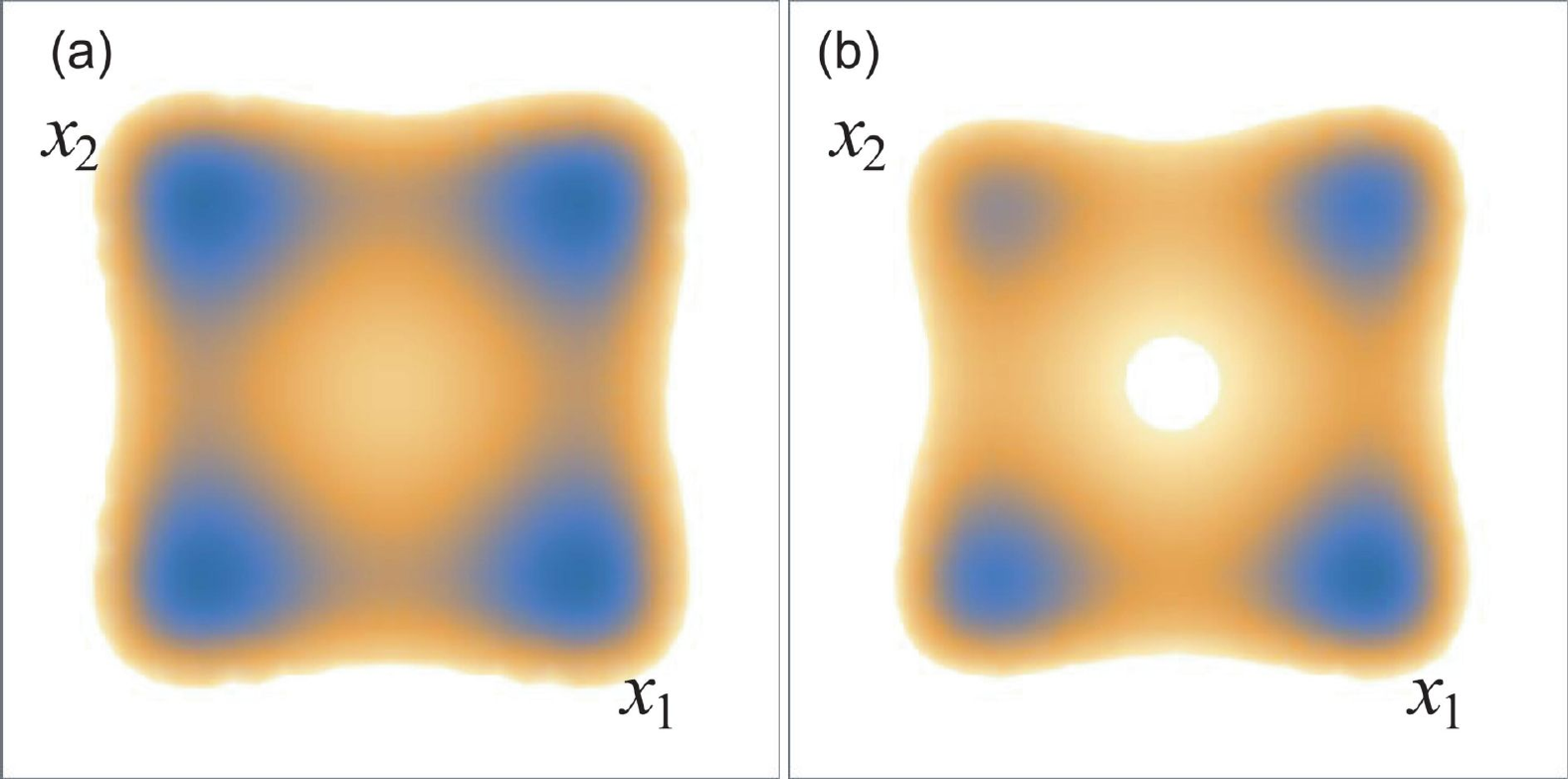}}
\caption{Density plot of energy spectrum (\protect\ref{TetraV}) in the $%
x_{1} $-$x_{2}$ plane. (a) Zero voltage $V_{1}=0$. (b) Nonzero voltage $%
\protect\varepsilon _{0}SV_{1}^{2}/2=100$. We have set $X_{\text{cap}}=10x_{%
\text{u}} $ and $a=2x_{\text{u}}$. }
\label{FigTetraWell}
\end{figure}

\section{Electrostatic energy}

In the parallel-plate NEMS, the capacitance between the plates $x_{1}$ and $%
x_{2}$ is well described by

\begin{equation}
C_{\text{para}}(x_{1},x_{2})=\frac{\varepsilon _{0}S}{X_{\text{cap}%
}+x_{1}-x_{2}},  \label{Cu}
\end{equation}%
where $X_{\text{cap}}$ is the distance between the adjacent plates without
buckling, $S$\ is the area of the plate, and $\varepsilon _{0}$ is the
permittivity. The electrostatic potential is given by%
\begin{equation}
U_{\text{cap}}=\frac{C_{\text{para}}(x_{1},x_{2})}{2}V_{1}^{2},  \label{Ucap}
\end{equation}%
when we control the voltage $V_{1}$\ between the plates.

We consider a set of two adjacent plates, where the potential energy is
given by%
\begin{equation}
V\left( x_{1},x_{2}\right) \equiv V\left( x_{1}\right) +V\left( x_{2}\right)
+\frac{C_{\text{para}}(x_{1},x_{2})}{2}V_{1}^{2}.  \label{TetraV}
\end{equation}%
We show the potential in the $x_{1}$-$x_{2}$ plane in Fig.\ref{FigTetraWell}%
. In the absence of the applied voltage $V_{1}=0$, there are four-fold
degenerated bottoms at $u_{1}=\pm a$ and $u_{2}=\pm a$ as shown in Fig.\ref%
{FigTetraWell}(a), where the ground state energy is $\hbar \omega $. Under
applied voltage, they are split as%
\begin{eqnarray}
V\left( a,a\right) &=&V\left( -a,-a\right) =\frac{\varepsilon _{0}S}{X_{%
\text{cap}}}\frac{V_{1}^{2}}{2}\equiv E_{0}, \\
V\left( a,-a\right) &=&\frac{\varepsilon _{0}S}{X_{\text{cap}}+2a}\frac{%
V_{1}^{2}}{2}\equiv E_{+}, \\
V\left( -a,a\right) &=&\frac{\varepsilon _{0}S}{X_{\text{cap}}-2a}\frac{%
V_{1}^{2}}{2}\equiv E_{-}.
\end{eqnarray}%
We also plot the potential profile with nonzero voltage in Fig.\ref%
{FigTetraWell}(b). We assume that the plate distance $X_{\text{cap}}$ is
very large compared with the deviation $a$ and obtain%
\begin{equation}
E_{+}-E_{0}\simeq -\frac{a}{X_{\text{cap}}^{2}}\varepsilon
_{0}SV_{1}^{2},\qquad E_{-}-E_{0}\simeq \frac{a}{X_{\text{cap}}^{2}}%
\varepsilon _{0}SV_{1}^{2}.  \label{Eapprox}
\end{equation}

\section{CNOT gates}

The CNOT gate $U_{\text{CNOT}}^{1\rightarrow 2}$ is defined by%
\begin{equation}
U_{\text{CNOT}}^{1\rightarrow 2}\equiv \left( 
\begin{array}{cccc}
1 & 0 & 0 & 0 \\ 
0 & 1 & 0 & 0 \\ 
0 & 0 & 0 & 1 \\ 
0 & 0 & 1 & 0%
\end{array}%
\right) ,
\end{equation}%
where the first qubit is the controlled qubit and the second qubit is the
target qubit.

\end{document}